\documentclass[aps,prd,onecolumn,groupedaddress,showpacs,nofootinbib,amssymb]{revtex4-2}
\usepackage[dvips]{graphicx}
\usepackage{amssymb}
\usepackage{amsmath}
\usepackage{graphicx,,color}
\usepackage{amsfonts}
\usepackage{bm}
\usepackage{cancel}
\usepackage{comment}
\usepackage{floatflt}

\newcommand\be{\begin{equation}}
\newcommand\ee{\end{equation}}

\allowdisplaybreaks[4]

\begin{document}

\title{A Power-law Inflation Tail for the Standard $R^2$-Inflation and the Trans-Planckian Censorship Conjecture}
\author{S.D.~Odintsov$^{1,2}$}\email{odintsov@ice.csic.es}
\author{V.K. Oikonomou,$^{3,4}$}\email{voikonomou@gapps.auth.gr;v.k.oikonomou1979@gmail.com}
\affiliation{$^{1)}$ Institute of Space Sciences (ICE, CSIC) C. Can Magrans s/n, 08193 Barcelona, Spain \\
$^{2)}$ ICREA, Passeig Luis Companys, 23, 08010 Barcelona, Spain\\
$^{3)}$Department of Physics, Aristotle University of
Thessaloniki, Thessaloniki 54124, Greece\\
$^{4)}$L.N. Gumilyov Eurasian National University - Astana,
010008, Kazakhstan}


 \tolerance=5000

\begin{abstract}
The conceptual problems of the standard slow-roll inflationary
scenario include the Trans-Planckian Censorship Conjecture issue,
which severely restricts the tensor-to-scalar ratio in the
standard minimally coupled scalar field inflation. Motivated by
the fact that a scalar field in its vacuum configuration can be
minimally coupled to gravity, or conformally coupled, and also
that the first quantum corrections of the scalar field action
include $R^2$ corrections, in this work we assume that $R^2$
gravity in the presence of a scalar field with constant equation
of state parameter co-exist and control the early Universe.
Constant equation of state parameter scalar field result from
exponential scalar potentials. In our approach, the standard
slow-roll era is controlled by the $R^2$ gravity and is followed
by a power-law inflationary tail governed by a minimally coupled
scalar field with an quintessential equation of state parameter,
stemming from an exponential scalar potential. The fact that the
total equation of state parameter after the end of the slow-roll
era is equal to the value determined by the scalar field, has an
effect on the duration of the $R^2$ governed slow-roll era, and it
actually shortens the duration of the slow-roll era, by an extent
which depends on the reheating temperature too, after all the
inflationary patches have ended. The power-law inflationary tail
to the standard $R^2$ inflation, solves the Trans-Planckian
Censorship Conjecture issues, and also the Swampland conjecture
can be amended in this context. We also perform a dynamical system
study to confirm numerically our findings.
\end{abstract}


\maketitle

\section{Introduction and Conceptual Motivation}

One of the most fundamental theories for the primordial Universe
is inflation \cite{inflation1,inflation2,inflation3,inflation4}, a
theory which can actually be realistically tested and verified
observationally. Currently, inflation is being scrutinized
observationally by the Planck collaboration \cite{Planck:2018vyg},
but promising new experiments and observational collaboration are
expected to further probe the effects of the inflationary era.
Specifically, the stage 4 Cosmic Microwave Background (CMB)
experiments \cite{CMB-S4:2016ple,SimonsObservatory:2019qwx} will
directly probe the $B$-modes of the inflationary era, if these
exist, and this will be a smoking gun for the direct verification
of inflation, since the $B$-mode pattern in the CMB will reveal
the tensor perturbations. There is an other way of indirectly
verifying the occurrence of inflation, by detecting a stochastic
gravitational wave background in future gravitational wave
experiments like LISA, DECIGO, Einstein Telescope and so on
\cite{Hild:2010id,Baker:2019nia,Smith:2019wny,Crowder:2005nr,Smith:2016jqs,Seto:2001qf,Kawamura:2020pcg,Bull:2018lat,LISACosmologyWorkingGroup:2022jok}.
In 2023 NANOGrav and other Pulsar Timing Arrays experiments have
confirmed the existence of a stochastic gravitational wave
background
\cite{nanograv,Antoniadis:2023ott,Reardon:2023gzh,Xu:2023wog},
however inflation is at odds in explaining that stochastic signal
\cite{Vagnozzi:2023lwo}. In principle, if the future gravitational
wave experiments reveal a stochastic gravitational wave
background, it could be challenging to pinpoint which theory can
explain the stochastic signal. To this end, the synergistic
approach of multiple distinct experiments will be needed or even
different theories can explain the combined stochastic
gravitational wave background pattern \cite{Odintsov:2024grb}.

There exist various theories that may describe an inflationary
era, standard inflation is described by a minimally coupled scalar
field theory, in which the dominance of the potential during the
inflaton evolution is profound, known as slow-roll evolution,
which is needed to provide solution to the flatness problems of
standard Big Bang cosmology. Inflation also solves other known
problems of standard Big Bang cosmology, such as the horizon and
monopoles problems, see for example
\cite{inflation1,inflation2,inflation3,inflation4}. Apart from the
standard minimally coupled scalar field description of inflation,
there exists the possibility of having a geometrical induced
inflationary era, being generated by some modified gravity
\cite{reviews1,reviews2,reviews3,reviews4}. The geometric
description of inflation and also of other evolutionary eras of
our Universe is strongly motivated for various reasons, firstly
the standard slow-rolling scalar field description of inflation
requires multiple couplings of the inflaton to the Standard Model
particles in order for reheating to occur in the Universe, which
is rather unappealing and somewhat fine-tuned. Also, the general
relativistic approaches toward describing the dark energy era
cannot easily accommodate a phantom evolution, which is
observationally allowed, in a concise and formally rigid manner.
In both cases, modified gravity offers conceptually appealing
descriptions, thus it is strongly motivated for describing both
inflation and the dark energy era. The most important modified
gravity candidate theory, and the simplest one, is $F(R)$ gravity,
\cite{Nojiri:2003ft,Capozziello:2005ku,Hwang:2001pu,Cognola:2005de,Song:2006ej,Faulkner:2006ub,Olmo:2006eh,Sawicki:2007tf,Faraoni:2007yn,Carloni:2007yv,
Nojiri:2007as,Deruelle:2007pt,Appleby:2008tv,Dunsby:2010wg} in the
context of which, inflation and dark energy can be described in a
unified way, see the pioneer article on this \cite{Nojiri:2003ft}.
From a theoretical point of view, it is highly possible that some
modified gravity controls the primordial Universe, in the presence
of some scalar field. Scalar fields are motivated and argued to be
present in our classical four dimensional Universe by string
theory. Often, scalars or axion-like particle, are the moduli of
some fundamental string theory, so it is possible to find remnants
of these moduli in our ordinary four dimensional Universe. But how
does modified gravity is involved in this context? When a scalar
field, possibly a remnant from string theory, is evaluated in its
vacuum configuration, it can either be minimally coupled or
conformally coupled, and therefore the quantum corrections to the
action will either be included in conformally coupled single
scalar field Lagrangians, or alternatively to minimally coupled
scalar field Lagrangians. The quantum corrections include higher
order derivatives of the metric, in the form of $R^2$, $R^3$ or
coupled forms of the Ricci and Riemann tensors as we evince in the
next section. Thus it is highly possible that a combined
$f(R,\phi)$ theory controls the primordial Universe.

On the other hand, inflation in the context of a slow-rolling
scalar field faces quite strong challenges, related with the
Swampland criteria and also the Trans-Planckian Censorship
Conjecture (TCC). The latter results in severely constraining the
value of the scalar field potential at the end of inflation and
also the tensor-to-scalar ratio, thus putting the whole standard
inflationary framework into a consistency peril. These problems
can be dealt in various ways, see Ref. \cite{Kamali:2020drm} and
references therein, resorting to non-standard multi-inflationary
theories and so on. We shall adopt the approach used by Ref.
\cite{Kamali:2020drm} where instead of a string-gas cosmology
followed by a short scalar field inflationary era, we shall assume
a standard $R^2$ inflationary era followed by a short period of
inflationary power-law tail evolution driven by a constant
equation of state (EoS) scalar field. The inflationary power-law
tail driven by the scalar field is generated by a scalar field
having an exponential potential. Exponential potentials in
minimally coupled scalar fields are motivated by several factors
in cosmology, both theoretical and phenomenological. These
potentials arise in different contexts within high-energy physics
and cosmology, particularly in the study of inflation, dark
energy, and scalar field cosmology. In the next section we further
analyze these motivations for using exponential potentials in more
detail. In our context, the inflationary power-law tail era takes
place right after the slow-roll inflationary era governed by $R^2$
gravity, thus it is essentially part of the inflationary era. The
total EoS of the Universe will be taken to be $w_{eff}<-1/3$
during this inflationary power-law tail period, in effect the
total number of $e$-foldings of the slow-roll inflationary era is
somewhat shortened. Apart from this feature, the inflationary
power-law tail era resolves the Swampland issues and the TCC
issues, since the evolution of the Universe at the end of the
inflationary era is basically a inflationary power-law evolution.
We show this in some detail. We also extend our approach by
allowing an interaction between the scalar field effective fluid
and the dark matter fluid, which may dominate the evolution right
after the inflationary era and before the radiation domination era
commences after the inflationary power-law tail. We study the
induced autonomous dynamical system formed by the scalar fluid and
the dark matter fluid, in the presence and absence of an
interaction between them. The fixed points of the induced
dynamical system contain precious information regarding the
dynamics of the mixed fluid system and as we show, dark matter
plays no role in determining the attractors of the cosmological
system, which are controlled by the scalar field.

This article is outlined as follows: In section II we discuss the
motivation for using a combined $R^2$ gravity with a constant EoS
scalar field with exponential potential. We mention the problems
of standard cosmology with a minimally coupled scalar and we
motivate how the combined $R^2$-constant EoS-scalar field theory
is theoretically motivated. In section III we analyze the combined
$R^2$ inflation era followed by a scalar field dominated
inflationary power-law tail. We show how this evolutionary
combination may resolve the TCC and Swampland issues occurring is
standard slow-roll inflation. In section IV we demonstrate how the
inclusion of a inflationary power-law tail era at the end of
inflation results in a shortening of the $e$-foldings number for
the slow-roll inflationary era, and we investigate the changes in
the inflationary phenomenology imposed by this $e$-foldings number
shortening. In section V we study the dynamical system of fluids
composed by the dark matter fluid and the constant EoS scalar
field fluid, in the presence and absence of an interaction between
them. We analyze the phase space trajectories and we highlight the
physical significance of the fixed points of the system, in the
presence and absence of an interaction between them. Finally the
conclusions of the article are presented in the end of the
article.

\section{Motivation to Use an $f(R,\phi)$ Gravity Theoretical Framework for Inflation and Problems of Standard Inflation}

The standard inflationary paradigm of a minimally coupled scalar
field, has some serious issues related with the Swampland criteria
and the TCC. Specifically, the TCC, initially proposed in
\cite{Martin:2000xs,Brandenberger:2000wr,Bedroya:2019snp,Brandenberger:2021pzy,Brandenberger:2022pqo,Berera:1995ie}
states that scales which were Trans-Planckian during the early
Universe, must remain inside the Hubble horizon for all times and
never enter in the classical evolution of our Universe. Assuming
that this conjecture is correct, this imposes a stringent bound on
the upper value of the scalar field potential at the end of
inflation, namely $V_e<10^{10}\,$GeV, which in turn restricts the
tensor-to-scalar ratio to have an upper bound $r<10^{-30}$. It is
conceivable that such a bound basically eliminates most of the
candidate models of minimally coupled scalar field theories. In
this work we aim to present a way to reconcile an $R^2$
inflationary theory with the TCC bound, by adding a inflationary
power-law tail in it, driven by a scalar fluid component in the
Universe, in which the scalar field had an exponential potential
and therefore has a constant EoS parameter during its evolution.
In this section we shall discuss how such a combined scalar field
and $R^2$ theory is theoretically motivated, and why an
exponential model for the scalar field may be a correct and
motivated description for an early Universe scalar field
component.

To start off, let us consider first the motivation for using a
scalar field theory in the presence of an $R^2$ gravity term. Let
us consider a scalar field evaluated in its vacuum configuration,
so it can be minimally coupled or conformally coupled. Therefore,
the quantum corrections will be considered for either the
non-minimally or minimally coupled scalar field theories. To
quantify our considerations, the most general scalar field theory
Lagrangian in four dimensions and which contains, at most, order
two derivatives, is the following,
\begin{equation}\label{generalscalarfieldaction}
\mathcal{S}_{\varphi}=\int
\mathrm{d}^4x\sqrt{-g}\left(\frac{1}{2}Z(\varphi)g^{\mu
\nu}\partial_{\mu}\varphi
\partial_{\nu}\varphi+\mathcal{V}(\varphi)+h(\varphi)R
\right)\, ,
\end{equation}
and as we already mentioned, when the scalar field is evaluated in
its vacuum configuration, the scalar field must be conformally
coupled or minimally coupled. For the purposes of this paper we
focus on the minimally coupled scalar field theory so we take
$Z(\varphi)=-1$ and $h(\varphi)=1$ in the gravitational action
(\ref{generalscalarfieldaction}). The quantum corrected effective
action for the above gravitational action
(\ref{generalscalarfieldaction}), which is compatible with
diffeomorphism invariance and also contains up to fourth order
derivatives, have the following form \cite{Codello:2015mba},
\begin{align}\label{quantumaction}
&\mathcal{S}_{eff}=\int
\mathrm{d}^4x\sqrt{-g}\Big{(}\Lambda_1+\Lambda_2
R+\Lambda_3R^2+\Lambda_4 R_{\mu \nu}R^{\mu \nu}+\Lambda_5 R_{\mu
\nu \alpha \beta}R^{\mu \nu \alpha \beta}+\Lambda_6 \square R\\
\notag & +\Lambda_7R\square R+\Lambda_8 R_{\mu \nu}\square R^{\mu
\nu}+\Lambda_9R^3+\mathcal{O}(\partial^8)+...\Big{)}\, ,
\end{align}
with the parameters $\Lambda_i$, $i=1,2,...,6$ being appropriate
dimensionful constants. Thus the combination of a scalar field and
of higher powers of the Ricci scalar is a rather natural selection
motivated theoretically by the one-loop corrected gravitational
action for a minimally coupled scalar field theory. In this work
we shall consider only the $R^2$ corrections, so in our case we
essentially have an $f(R,\varphi)$ gravitational action of the
form,
\begin{equation}
\label{action} \centering
\mathcal{S}=\int{d^4x\sqrt{-g}\left(\frac{
F(R)}{2\kappa^2}-\frac{1}{2}g^{\mu \nu}\partial_{\mu}\varphi
\partial_{\nu}\varphi-\mathcal{V}(\varphi)\right)}\, ,
\end{equation}
with $\kappa^2=8\pi G=\frac{1}{Mp^2}$ and also $M_p$ denotes the
reduced Planck mass, and in addition,
\begin{equation}\label{fr}
    F(R)=R+\frac{R^2}{M^2}\, ,
\end{equation}
where $M$ being a mass scale which is constrained by inflationary
phenomenology. In the literature, such combinations of
$R^2$-corrected scalar field theories have been considered in
\cite{Ema:2017rqn,Ema:2020evi,Ivanov:2021ily,Gottlober:1993hp,Enckell:2018uic,Kubo:2020fdd,Gorbunov:2018llf,Calmet:2016fsr,Oikonomou:2021msx,Oikonomou:2022bqb},
but in the context of inflation and without assuming that the
scalar field has a constant EoS parameter. In addition, in most of
the cases, the inflationary framework is considered in the
Einstein frame, thus treating the theory as a two scalar field
theory. Our approach is different since we consider the Jordan
frame theory, in which the perturbations that generate the large
scale structure of the Universe are generated during the $R^2$
inflationary era, and then the inflationary power-law tail driven
by the scalar field occurs right after the $R^2$ inflationary era.
In this inflationary era which is the tail of the slow-roll $R^2$
driven inflationary era, cosmological perturbations are also
generated but the modes that exit the horizon after the slow-roll
era have extremely small wavelength, certainly much smaller than
10Mpc, so these modes reenter the Hubble horizon at the first
stages of the radiation era, thus do not contribute to the linear
component of the CMB.

Now let us discuss the motivation to use a scalar field theory
with an exponential potential, which makes the scalar field have a
constant EoS parameter. Exponential potentials in minimally
coupled scalar fields are motivated by several factors in
cosmology, both theoretical and phenomenological. These
exponential potentials arise in different contexts within
high-energy physics and cosmology, particularly in the study of
inflation, dark energy, and scalar field cosmology. In certain
string theory models and higher-dimensional theories like
supergravity or Kaluza-Klein compactifications, exponential
potentials often appear naturally. For example, when fields like
moduli (parameters describing the size and shape of extra
dimensions) evolve, their dynamics can lead to effective
four-dimensional theories with exponential potentials for scalar
fields. The dimensional reduction from higher-dimensional theories
can also produce terms that behave like exponential potentials.
These often appear in the context of dilaton fields or other
axion-like particles associated with compactified extra
dimensions. Also exponential potentials are frequently used in
scalar-tensor theories and quintessence theories. Indeed, in
scalar-tensor theories of gravity, scalar fields that are
minimally coupled to gravity (but not directly to matter) can be
responsible for the late-time acceleration of the Universe
(quintessence). Exponential potentials are a common choice in
these models because they lead to attractor solutions, where the
scalar field evolves in a manner that mimics or dominates the
cosmic expansion. Quintessence models with exponential potentials
have a wide range of dynamical behaviors, including scaling
solutions where the energy density of the scalar field tracks that
of the dominant component (e.g., matter or radiation), which is
useful in explaining dark energy and cosmic acceleration. Also
attractor solutions are quite often related with exponential
potentials. This is one of the key motivations for using
exponential potential, namely the fact that these potentials lead
to attractor solutions in the equations of motion for scalar
fields. In these solutions, the scalar field evolves in a
predictable manner regardless of initial conditions. For example,
in inflationary models or in the late Universe (quintessence), an
exponential potential can lead to power-law inflation or scaling
behavior. Also, another motivation for using exponential
potentials, is scale-invariance and other symmetry arguments.
Exponential potentials can be motivated by considerations of scale
invariance or certain symmetries in field theory. The exponential
form respects scale invariance in some cases, which is a desirable
property for constructing models that are not fine-tuned or that
have certain symmetries in the early Universe. In certain
supersymmetric and non-supersymmetric string theory
compactifications, these potentials also preserve specific
symmetries related to the evolution of the moduli fields.

\section{How a Scalar Field Power-law Inflationary Tail of a Slow-roll Inflation can Reconcile Inflation with TCC}

Let us discuss now how a stiff era generated by a scalar field
with an exponential potential can reconcile inflation with the
TCC. We shall follow closely the considerations of
\cite{Kamali:2020drm} which assumed that an inflationary era with
constant EoS followed a string gas era. Our assumption is that a
inflationary power-law tail era follows a standard slow-roll $R^2$
inflationary era. If the scalar field theory has an exponential
potential, then the following condition holds necessarily true,
\begin{equation}\label{eoscondition}
\dot{\phi}^2=\beta V(\phi)\, ,
\end{equation}
thus,
\begin{equation}\label{dddotphi}
\ddot{\phi}=\frac{\beta V'}{2}\, .
\end{equation}
In Ref. \cite{Kamali:2020drm} the requirement was that $\beta<1$
in order to describe an inflationary power-law era, and we will
take $\beta=0.99$. The scalar field equation reads,
\begin{equation}\label{fieldfree}
\ddot{\phi}+3H\dot{\phi}+V'=0\, ,
\end{equation}
so in view of the above, we have,
\begin{equation}\label{eqnfreescalarkin}
\left(\frac{\beta+2}{2} \right)^2\left(
V'\right)^2=9H^2\dot{\phi}^2\, ,
\end{equation}
which in turn yields,
\begin{equation}\label{potentialapprox}
V=V_0e^{-\sqrt{\frac{6\beta}{\beta+2}}\kappa \phi}\, .
\end{equation}
This is exactly the potential we shall assume in this article, and
we define $\lambda$ as follows,
\begin{equation}\label{lambda}
\lambda=\sqrt{\frac{6\beta}{\beta+2}}\, ,
\end{equation}
so the scalar potential is written as follows,
\begin{equation}\label{exponentialpotential}
V(\phi)=V_0\,e^{-\lambda \phi \kappa}\, .
\end{equation}
The EoS parameter for the scalar field is equal to,
\begin{equation}\label{eosscalardef}
w_{\phi}=\frac{\frac{\dot{\phi}^2}{2}-V}{\frac{\dot{\phi}^2}{2}+V}\,
,
\end{equation}
hence for the exponential potential and due to the relation
(\ref{eoscondition}), we have,
\begin{equation}\label{eosscalarfinal}
w_{\phi}=\frac{\beta-2}{\beta+2}\, .
\end{equation}
Let us now consider how the inflationary power-law tail of a
standard $R^2$ slow-roll era may relax the TCC constraints on
standard inflation. Following closely the argument of Ref.
\cite{Kamali:2020drm}, applied in our case, the evolution of the
energy density of the scalar field after the end of the slow-roll
inflation has the following form,
\begin{equation}\label{b1}
\rho_{\phi}\sim a^{-\frac{3\beta}{1+\frac{\beta}{2}}}\, .
\end{equation}
In order to have spatial flatness in the Universe, the fractional
contribution of the spatial curvature to the critical energy
density $\Omega_K$ must satisfy,
\begin{equation}\label{b2}
\Omega_K<10^{-2}\frac{T_0T_{eq}}{T_R^2}\, ,
\end{equation}
where $T_0$, $T_{eq}$ and $T_R$ are the temperatures of the
Universe at present day, at matter-radiation equilibrium and at
the end of the inflationary power-law tail era, where we also
assumed that after the inflationary power-law tail era, the
Universe is directly radiation dominated. Demanding that the
degrease of the curvature density during the inflationary
power-law tail era is larger than the relative increase occurring
after the inflationary power-law tail era, we must have,
\begin{equation}\label{b3}
\left(\frac{a_i}{a_R}\right)^{2-\frac{3\beta}{1+\frac{\beta}{2}}}<10^{-2}\frac{T_0T_{eq}}{T_R^2}\,
,
\end{equation}
thus we assumed that the flatness issue of the Universe is already
resolved between the beginning of the inflationary power-law tail
era, with scale factor $a_i$ and the end of the inflationary
power-law tail era with scale factor $a_R$. We require that the
comoving scale which corresponds to the current Hubble radius is
larger than the Hubble length at the beginning of the inflationary
power-law tail era $H^{-1}(t_i)$, that is,
\begin{equation}\label{b4}
H^{-1}(t_i)<H_0^{-1}\frac{T_0 a_i}{T_Ra_R}\, .
\end{equation}
During the inflationary power-law tail era, the Friedmann equation
is,
\begin{equation}\label{b5}
3H^2=\kappa^2\left(1+\frac{\beta}{2} \right)V\, ,
\end{equation}
and also we currently have,
\begin{equation}\label{b6}
H_0^2=\frac{\kappa^2}{3}g^{*}T_0^3T_{eq}\, ,
\end{equation}
with $g^*$ being the number of relativistic degrees of freedom. If
a radiation domination era follows after the inflationary
power-law tail era, we have,
\begin{equation}\label{b7}
\left( 1+\frac{\beta}{2}\right)V_R=g^*T_R^4\, ,
\end{equation}
where $T_R$ and $V_R$ are the temperature and the potential at the
end of the inflationary power-law tail era. Thus we have,
\begin{equation}\label{b8}
\frac{V_R}{V_i}<\frac{T_R^2}{T_0T_{eq}}\left(\frac{a_i}{a_R}
\right)^2\, ,
\end{equation}
where $V_i$ is the scalar potential value at the beginning of the
inflationary power-law tail era. So by using the Friedmann
equations, which for the exponential potential yield a power-law
evolution for the scale factor, we have,
\begin{equation}\label{b9}
\frac{V_i}{V_R}=\left( \frac{a_R}{a_i}
\right)^{\frac{3\beta}{\sqrt{1+\frac{\beta}{2}}}}\, ,
\end{equation}
and then Eq. (\ref{b8}), becomes,
\begin{equation}\label{b10}
\left( \frac{a_i}{a_R}
\right)^{2-\frac{3\beta}{\sqrt{1+\frac{\beta}{2}}}}>\frac{T_0T_{eq}}{T_R^2}\,
,
\end{equation}
If the inflationary power-law tail era was a period of slow-roll
inflation, in which $\beta\sim 0$ ($w_{\phi}\sim -1$), the above
inequality and that of Eq. (\ref{b3}) would be in conflict.
However, for large values of $\beta$, the two inequalities are
compatible. In this paper we shall consider $\beta\sim 0.99$ which
yields $w_{\phi}\sim -0.337793$, which is a slightly
quintessential acceleration EoS. Note however, that the scale
factor $a_i$ in both Eqs. (\ref{b3}) and (\ref{b10}) describes the
scale factor at the end of the preceding $R^2$ slow-roll era, and
at the same time the scale factor of the Universe at the beginning
of the power-law tail. Now in the hypothetical scenario that the
power-law tail was another slow-roll era, sequential to the $R^2$
slow-roll era, then Eqs. (\ref{b3}) and (\ref{b10}) would indeed
be in conflict. But the core assumption of this work is that the
power-law tail with initial scale factor $a_i$ is not a slow-roll
sequence of the $R^2$ slow-roll era, but a power-law evolution
with $\beta=0.99$. Thus for this value of $\beta$, Eqs. (\ref{b3})
and (\ref{b10}) are rendered compatible.

Regarding the TCC, this is expressed mathematically as,
\begin{equation}\label{tcc}
\frac{a_R}{a_i}\ell_{pl}<H^{-1}(t_R)\, ,
\end{equation}
where $\ell_{pl}$ is the Planck length. Following closely
\cite{Kamali:2020drm}, this can be written,
\begin{equation}\label{eqextra}
\left( \frac{a_i}{a_R}\right)^2>\frac{g^*}{3}\left(\frac{T_R}{M_p}
\right)^4\, .
\end{equation}
In order for inflation to solve the flatness problem, there is the
bound of Eq. (\ref{b3}), which can be consistent with the TCC only
if the following inequality is satisfied by the temperature at the
end of the inflationary power-law tail of standard slow-roll $R^2$
inflation, which is basically the energy scale at the end of
inflation \cite{Kamali:2020drm},
\begin{equation}\label{tr}
\frac{T_R}{M_p}<\left( 3/
g^*\right)^{(1-\tilde{\beta})/(6-4\tilde{\beta})}\times10^{-2/(6-4\tilde{\beta})}\left(\frac{T_0T_{eq}}{M_p}
\right)^{1/(6-4\tilde{\beta})}\, ,
\end{equation}
where $\tilde{\beta}=\frac{3\beta}{\beta+2}$. For our case, for
$\beta=0.99$ we must have a significantly small temperature after
the end of the power-law inflationary tail era $T_R<10^{-30}\times
M_p$, however the flatness issue is already resolved by the $R^2$
inflation era if it lasts for at least $\geq 48$ $e$-folds in the
worst case, as we show later on. Also such a low final temperature
of the Universe at the end of the power-law tail, is not in
conflict with the tensor-to-scalar ratio or the scalar
perturbations, since the power-law inflationary patch is
completely detached from the slow-roll patch, and does not
contribute at all to the CMB. In Ref. \cite{Kamali:2020drm}, this
final inflationary temperature was found to be of the order $\sim
240\,$GeV, since the authors used a value $\beta=1/2$. We used
$\beta=0.99$ which basically means that the inflationary power-law
scalar field driven patch is nearly quintessential (the value
$\beta=0.99$ yields an EoS parameter slightly smaller than the
non-acceleration value $w=-1/3$). We need to note that if someone
uses a larger $\beta$, then the temperature at the end of the
inflationary power-law tail of $R^2$ inflation can increase and
become significantly larger, depending on the value of $\beta$.
But when $\beta>1$ we are no longer talking about an inflationary
patch, but for a scalar field dominated evolution with EoS
$-1/3<w\leq 1$, but this scenario has a lot of different features,
which could be studied in another work.

Finally regarding the Swampland constraint,
\begin{equation}\label{swamp}
\frac{|V'|}{V}\geq \frac{c}{M_p}\,
\end{equation}
where $c\sim \mathcal{O}(1)$, and in our case, taking into account
that it applies during the power-law inflationary era, we have,
\begin{equation}\label{swamp}
\frac{|V'|}{V}=\frac{|-\lambda|}{M_p}\, ,
\end{equation}
and since for $\beta\sim 0.99$ we have $\lambda\sim 1.40$, the
Swampland criterion is satisfied. Thus in this section we
demonstrated that a inflationary power-law tail extension of a
standard inflationary era may smoothen the standard problems of
canonical slow-roll inflation. In the next section we also
demonstrate that the duration of the slow-roll inflation era is
also shortened if the total EoS at the end of inflation is
distinct from that of radiation.

Now an important comment is in order regarding whether the
sub-Planckian modes exit the Hubble horizon during the $R^2$
inflation era. When the $R^2$ inflation begins at first horizon
crossing, the Hubble horizon is quite large and gradually shrinks
in an inverse exponential way, since the $R^2$ model produces a
nearly de Sitter evolution. Thus during the $R^2$ inflation, the
sub-Planckian modes remain way too deeply in the Hubble horizon.
As $R^2$ inflation proceeds though, more and more low wavelength
modes exit the horizon. Thus the peril of having sub-Planckian
modes exiting the Hubble horizon increases. Thus the issue with
sub-Planckian modes exiting the horizon concerns the last stages
of the $R^2$ inflationary era, where the Hubble horizon has
significantly shrunk and may include the sub-Planckian modes. Thus
the problem with sub-Planckian modes may occur at the last
e-foldings of inflation where the Hubble horizon is too small and
may include some of the sub-Planckian modes. However, in our
analysis this is where the power-law tail occurs, thus the
sub-Planckian modes remain sub-Hubble with this power-law tail
continuation of the $R^2$ inflation era.

\section{Inflation in $F(R)$ Gravity in the Presence of a Scalar Field}

Let us consider now the inflationary phenomenology of the
gravitational action (\ref{action}) with the $F(R)$ gravity being
the $R^2$ model of Eq. (\ref{fr}). The parameter $M$ appearing in
Eq. (\ref{action}) is chosen to be $M= 1.5\times
10^{-5}\left(\frac{N}{50}\right)^{-1}M_p$, for phenomenological
reasoning \cite{Appleby:2009uf}, and $N$ denotes the $e$-foldings
number. For a flat Friedmann-Robertson-Walker (FRW) metric,
\begin{equation}
\label{metricfrw} ds^2 = - dt^2 + a(t)^2 \sum_{i=1,2,3}
\left(dx^i\right)^2\, ,
\end{equation}
the field equations read,
\begin{align}\label{eqnsofmkotion}
& 3 H^2F_R=\frac{RF_R-F}{2}-3H\dot{F}_R+\kappa^2\left(
\rho_r+\frac{1}{2}\dot{\phi}^2+V(\phi)\right)\, ,\\ \notag &
-2\dot{H}F=\kappa^2\dot{\phi}^2+\ddot{F}_R-H\dot{F}_R
+\frac{4\kappa^2}{3}\rho_r\, ,
\end{align}
\begin{equation}\label{scalareqnofmotion}
\ddot{\phi}+3H\dot{\phi}+V'(\phi)=0
\end{equation}
with $F_R=\frac{\partial F}{\partial R}$, and the ``dot''
indicates a differentiation with respect to $t$, while the
``prime'' denotes differentiation with respect to $\phi$. For our
considerations we shall assume that the slow-roll era has a low
inflationary scale of the order $H_I=10^{13}$GeV, and in addition
we further assume that  $V_0$ in Eq. (\ref{potentialapprox}) of
the scalar field potential is slightly smaller than the Planck
allowed value due to the amplitude of the scalar perturbations,
and specifically we assume that $V_0\sim 9.6\times 10^{-15}\times
M_p^4$. Hence, for a sub-Planckian valued scalar field, the
potential has values of the order $\kappa^2 V\sim 1.44935\times
10^{42}\,$eV$^2$, where we took $\beta\sim 0.99$. Now, for
$H_I=10^{13}$GeV and $N\sim 60$, the parameter $M$ is
approximately $M\simeq 3.04375\times 10^{22}$eV and assuming a
slow-roll era initially, we have $R\sim 1.2\times 10^{45}$eV$^2$
and also $R^2/M^2\sim \mathcal{O}(1.55\times 10^{45})$eV$^2$.
Apparently, due to the constant EoS evolution of the scalar field
controlled by Eq. (\ref{eoscondition}), the scalar field kinetic
terms are of the same order as the potential, thus the $R^2$
gravity term dominates initially the inflationary era, which
recall is a slow-roll era. Hence, the field equations initially
acquire the form,
\begin{equation}\label{patsunappendix}
\ddot{H}-\frac{\dot{H}^2}{2H}+\frac{H\,M^2}{2}=-3H\dot{H}\, .
\end{equation}
and since the slow-roll conditions apply, we get,
\begin{equation}\label{patsunappendix1}
-\frac{M^2}{6}=\dot{H}\, ,
\end{equation}
which has a quasi-de Sitter evolution as solution,
\begin{equation}\label{quasidesitter}
H(t)=H_I-\frac{M^2}{6} t\, .
\end{equation}
The inflationary evolution of $F(R)$ gravity is quantified by the
dynamics of the slow-roll indices,
\cite{Hwang:2005hb,reviews1,Odintsov:2020thl},
\begin{equation}
\label{restofparametersfr}\epsilon_1=-\frac{\dot{H}}{H^2}, \quad
\epsilon_2=0\, ,\quad \epsilon_3= \frac{\dot{F}_R}{2HF_R}\, ,\quad
\epsilon_4=\frac{\ddot{F}_R}{H\dot{F}_R}\,
 ,
\end{equation}
and the spectral index of the primordial scalar perturbations and
the tensor-to-scalar ratio are written as follows
\cite{reviews1,Hwang:2005hb},
\begin{equation}
\label{epsilonall} n_s=
1-\frac{4\epsilon_1-2\epsilon_3+2\epsilon_4}{1-\epsilon_1},\quad
r=48\frac{\epsilon_3^2}{(1+\epsilon_3)^2}\, .
\end{equation}
Using the Raychaudhuri equation for $F(R)$ gravity, we obtain,
\begin{equation}\label{approx1}
\epsilon_1=-\epsilon_3(1-\epsilon_4)\, ,
\end{equation}
therefore we have approximately,
\begin{equation}
\label{spectralfinal} n_s\simeq 1-6\epsilon_1-2\epsilon_4\, ,
\end{equation}
and also,
\begin{equation}
\label{tensorfinal} r\simeq 48\epsilon_1^2\, .
\end{equation}
Also considering $\epsilon_4=\frac{\ddot{F}_R}{H\dot{F}_R}$ we
get,
\begin{equation}\label{epsilon41}
\epsilon_4=\frac{\ddot{F}_R}{H\dot{F}_R}=\frac{\frac{d}{d
t}\left(F_{RR}\dot{R}\right)}{HF_{RR}\dot{R}}=\frac{F_{RRR}\dot{R}^2+F_{RR}\frac{d
(\dot{R})}{d t}}{HF_{RR}\dot{R}}\, ,
\end{equation}
and due to the fact that,
\begin{equation}\label{rdot}
\dot{R}=24\dot{H}H+6\ddot{H}\simeq 24H\dot{H}=-24H^3\epsilon_1\, ,
\end{equation}
combined with Eq. (\ref{epsilon41}) we get,
\begin{equation}\label{epsilon4final}
\epsilon_4\simeq -\frac{24
F_{RRR}H^2}{F_{RR}}\epsilon_1-3\epsilon_1+\frac{\dot{\epsilon}_1}{H\epsilon_1}\,
.
\end{equation}
By using,
\begin{equation}\label{epsilon1newfiles}
\dot{\epsilon}_1=-\frac{\ddot{H}H^2-2\dot{H}^2H}{H^4}=-\frac{\ddot{H}}{H^2}+\frac{2\dot{H}^2}{H^3}\simeq
2H \epsilon_1^2\, ,
\end{equation}
$\epsilon_4$ becomes,
\begin{equation}\label{finalapproxepsilon4}
\epsilon_4\simeq -\frac{24
F_{RRR}H^2}{F_{RR}}\epsilon_1-\epsilon_1\, .
\end{equation}
The spectral index of the scalar perturbations is,
\begin{equation}\label{scalarspectralindex}
n_{\mathcal{S}}=1-4\epsilon_1-2\epsilon_2+2\epsilon_3-2\epsilon_4\,
,
\end{equation}
which in view of the above equations, is simplified as follows,
\begin{equation}\label{spectralindexfinalform}
n_{\mathcal{S}}\simeq 1-(2-x)\epsilon_1+2\epsilon_3\, .
\end{equation}
In addition, the scalar-to-tensor ratio in terms of the slow-roll
indices is \cite{Hwang:2005hb,reviews1,Odintsov:2020thl},
\begin{equation}\label{tensortoscaalrratio}
r\simeq 48\epsilon_1^2\, .
\end{equation}
For the $R^2$ gravity, $x=0$, therefore we have,
\begin{equation}\label{epsilon1indexanalytic}
\epsilon_1=-\frac{6 M^2}{\left(M^2 t-6 H_I\right)^2}\, ,
\end{equation}
and also by solving$\epsilon_1(t_f)=1$, we obtain the time
instance $t_f$ where inflation ends,
\begin{equation}\label{finaltimeinstance}
t_f=(6 H_I + \sqrt{6} M)/M^2\, .
\end{equation}
Using,
\begin{equation}\label{efoldingsnumber}
N=\int_{t_i}^{t_f}H(t)dt\, ,
\end{equation}
and also in conjunction with Eq. (\ref{finaltimeinstance}), the
time instance of the first horizon crossing is,
\begin{equation}\label{ti}
t_i=\frac{2 \sqrt{9 H_I^2-3 M^2 Y}+6 H_I}{M^2}\, ,
\end{equation}
hence the first slow-roll index at the first horizon crossing
reads,
\begin{equation}\label{epsilon1lambdaind}
\epsilon_1(t_i)=\frac{1}{1+2N}\, ,
\end{equation}
therefore the spectral index and the tensor-to-scalar ratio
acquire the following forms $n_s\sim 1-\frac{2}{N}$ and $r\sim
\frac{12}{N^2}$. The dominance of the $R^2$ term ceases to occur
after nearly $60$ $e$-folds, where $\epsilon_1$ approaches unity.
After that the scalar field starts to dominate the evolution and
the power-law inflationary tail takes place.

\begin{figure}[h!]
\centering
\includegraphics[width=20pc]{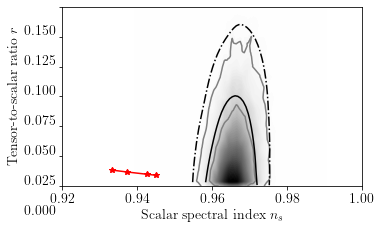}
\includegraphics[width=20pc]{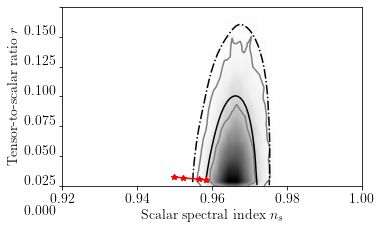}
\includegraphics[width=20pc]{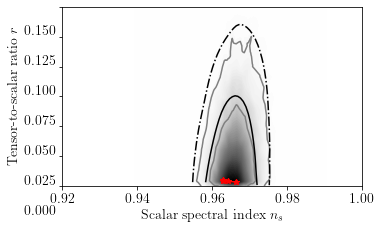}
\caption{Inflationary phenomenology for $\beta=0.99$, and
$w=-0.337793$ versus the Planck 2018 data. In all the plots, the
black thick line corresponds to the Planck 2018 likelihood curve
at $95\%$CL, the dash dotted curve to the Planck 2018 likelihood
curve at $68\%$CL. In the upper left plot, the red curve
corresponds to $T_r=10^2\,$GeV for $N=[30,36]$, in the upper right
plot the red curve corresponds to $T_r=10^7\,$GeV for $N=[40,48]$
and in the bottom plot the red curve corresponds to
$T_r=10^{12}\,$GeV for $N=[50,59]$. As it can be seen in this case
too, only the high reheating temperature phenomenology is viable.}
\label{p2rehe}
\end{figure}

An important comment is in order. There is the question whether
the power-law inflationary tail produces primordial curvature
perturbations. The answer is of course yes, however, the
wavelength of these perturbations is particularly small, since
recall these modes exit the horizon during the power-law tail,
after the slow-roll era, hence, for sure, are not expected to
contribute to the linear modes of the CMB, probed by Planck data.
In fact, these modes that exit the Hubble horizon during the
power-law inflationary tail, will have probably a wavelength
$\lambda \ll 10\,$Mpc, thus these will re-enter the Hubble horizon
during the first steps of reheating and do not contribute to the
linear part of the CMB probed by the Planck collaboration.
However, due to the fact that after the slow-roll era, the
background EoS is not that of radiation, but is nearly $w\preceq
-1/3$, the total duration of the inflationary era is actually
shortened in the following way,
\cite{Adshead:2010mc,Munoz:2014eqa,Liddle:2003as},
\begin{equation}\label{efoldingsmainrelation}
N=56.12-\ln \left( \frac{k}{k_*}\right)+\frac{1}{3(1+w)}\ln \left(
\frac{2}{3}\right)+\ln \left(
\frac{\rho_k^{1/4}}{\rho_{end}^{1/4}}\right)+\frac{1-3w}{3(1+w)}\ln
\left( \frac{\rho_{reh}^{1/4}}{\rho_{end}^{1/4}}\right)+\ln \left(
\frac{\rho_k^{1/4}}{10^{16}\mathrm{GeV}}\right)\, ,
\end{equation}
where $k_*=0.05$Mpc$^{-1}$, is the CMB pivot scale, with $\rho_k$
standing for the total energy density of the Universe during the
first horizon crossing of the mode $k$ (primordially during the
commence of the slow-roll inflationary era), $\rho_{end}$ denotes
the Universe's energy density at the end of inflation, and also
$\rho_{reh}$ denotes the Universe's energy density at exactly the
end of the reheating era, with a reheating temperature $T_r$. An
important comment is in order at this point. With $T_r$, we denote
the reheating temperature, which is different from $T_R$, which is
the temperature of the Universe at the end of the power-law tail
and at the beginning of the reheating era, which appears in the
previous section in the inequality (\ref{tr}). So $T_r$ and $T_R$
are two different and distinct quantities. As we mentioned, $T_r$
is the reheating temperature, that is, the maximum temperature
that the Universe acquires, and it is reached at the end of the
reheating era, via some reheating mechanism, either geometrically
due to the $R^2$ theory and curvature perturbations, or due to
oscillations of the scalar field, or even both mechanisms. Now,
$T_R$ is the temperature of the Universe at the end of the
power-law tail, and at the onset of the reheating era.
\begin{table}[h!]
  \begin{center}
    \caption{\emph{\textbf{Maximum $e$-foldings number, and the inflationary observational indices for $\beta=0.99$, and $w=-0.337793$ for three reheating temperatures $T_r$.}}}
    \label{tablerehe1}
    \begin{tabular}{|r|r|r|r|}
     \hline
      \textbf{Reheating Temperature}   & \textbf{Maximum $e$-foldings} & \textbf{Spectral Index $n_s$} &
      \textbf{Tensor-to-scalar Ratio}
      \\  \hline
      $T_r=10^{2}$GeV  & $N_{max}=36.4299$ & $n_s=0.945$ & $r=0.009$
\\  \hline
 $T_r=10^{7}$GeV & $N_{max}=48$ & $n_s=0.9584$ & $r=0.00518$
\\  \hline
 $T_r=10^{12}$GeV & $N_{max}=59$ & $n_s=0.96653$ & $r=0.00335$
\\  \hline
    \end{tabular}
  \end{center}
\end{table}
In our case, for $\beta=0.99$, the value of the temperature at the
onset of the reheating era was found to be $T_R<10^{-30}\times
M_p$, just below Eq. (\ref{tr}) in the text. So $T_R\sim
10^{-11}\,$GeV hence, the Universe after the power-law tail and at
the beginning of the reheating era is quite cold. Now $T_r$ is the
reheating temperature, thus the maximum temperature reached during
the reheating era of the Universe. The reheating temperature $T_r$
is quite larger than $T_R$, see the choices for $T_r$ later on in
this section, where we shall use a high reheating temperature
$T_r=10^{12}$GeV, an intermediate $T_r=10^{7}$GeV, and a low
reheating temperature, $T_r=10^{2}$GeV.
\begin{table}[h!]
  \begin{center}
    \caption{\emph{\textbf{Maximum $e$-foldings number, and the inflationary observational indices for $\beta=0.9$, and $w=-0.37931$ for three reheating temperatures $T_r$.}}}
    \label{tablerehe2}
    \begin{tabular}{|r|r|r|r|}
     \hline
      \textbf{Reheating Temperature}   & \textbf{Maximum $e$-foldings} & \textbf{Spectral Index $n_s$} &
      \textbf{Tensor-to-scalar Ratio}
      \\  \hline
      $T_r=10^{2}$GeV  & $N_{max}=32$ & $n_s=0.93882$ & $r=0.011225$
\\  \hline
 $T_r=10^{7}$GeV & $N_{max}=45$ & $n_s=0.9564$ & $r=0.0056$
\\  \hline
 $T_r=10^{12}$GeV & $N_{max}=59$ & $n_s=0.96617$ & $r=0.00343$
\\  \hline
    \end{tabular}
  \end{center}
\end{table}
So these are quite larger than the temperature $T_R$ at the
beginning of the reheating era, or equivalently the end of the
power-law tail era. Also, one crucial assumption we make is that
we assume that the EoS of the Universe will not change during the
era after the end of the power-law tail, until the reheating
temperature is reached. Having this issue clarified, using
$\rho=\frac{\pi^2}{30}g_*T^4$ we can express the $e$-foldings
number in terms of the reheating temperature and other crucial
temperatures. Let us assume for simplicity three distinct
reheating temperatures, a high reheating temperature
$T_r=10^{12}$GeV, an intermediate $T_r=10^{7}$GeV, and a low
reheating temperature, $T_r=10^{2}$GeV. Although low-reheating
temperatures might seem unnatural, these occur frequently in the
literature \cite{Hasegawa:2019jsa}. In order to have a direct idea
of the shortening of the inflationary era due to the non-trivial
EoS after the slow-roll era, let us evaluate the $e$-foldings
number for the three distinct reheating temperatures and for
$w=-0.337793$, a value which corresponds to the choice
$\beta=0.99$. For $w=-0.337793$ and a reheating temperature
$T_r=10^{12}$GeV, we have, $N=59.7659$, while for $w=1/3$ we would
have $N=64.5358$. Thus the observational indices are in this case
marginally different than the radiation case, and specifically for
$w=-0.337793$ we have $n_{s}=0.966536$, $r=0.0033595$, while for
$w=1/3$ we have $n_{s}=0.969009$, $r=0.00288124$. Now for
$w=-0.337793$ and a reheating temperature $T_r=10^{7}$GeV, we
have, $N=48.0979$, while for $w=1/3$ we would have $N=64.5358$.
Thus the observational indices are in this case significantly
different than the radiation case, and specifically for
$w=-0.337793$ we have $n_{s}=0.958606$, $r=0.00514035$, while for
$w=1/3$ we have $n_{s}=0.969009$, $r=0.00288124$. Also for
$w=-0.337793$ and a reheating temperature $T_r=10^{2}$GeV, we
have, $N=36.4299$, while for $w=1/3$ we would have $N=64.5358$.
Thus the observational indices are in this case significantly
different than the radiation case, and specifically for
$w=-0.337793$ we have $n_{s}=0.9451$, $r=0.00904201$, while for
$w=1/3$ we have $n_{s}=0.969009$, $r=0.00288124$.
\begin{figure}[h!]
\centering
\includegraphics[width=20pc]{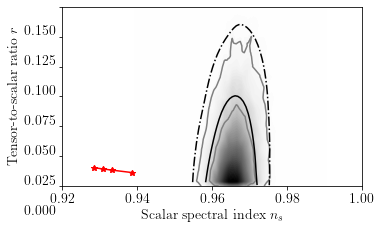}
\includegraphics[width=20pc]{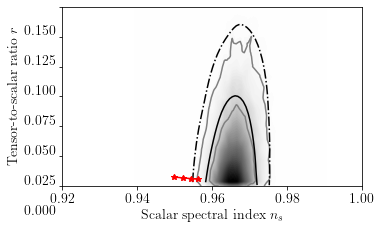}
\includegraphics[width=20pc]{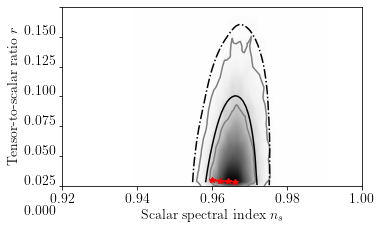}
\caption{Inflationary phenomenology for $\beta=0.9$, and
$w=-0.37931$ versus the Planck 2018 data. In all the plots, the
black thick line corresponds to the Planck 2018 likelihood curve
at $95\%$CL, the dash dotted curve to the Planck 2018 likelihood
curve at $68\%$CL. In the upper left plot, the red curve
corresponds to $T_r=10^2\,$GeV for $N=[28,32]$, in the upper right
plot the red curve corresponds to $T_r=10^7\,$GeV for $N=[40,45]$
and in the bottom plot the red curve corresponds to
$T_r=10^{12}\,$GeV for $N=[50,59]$. As it can be seen, only the
high reheating temperature phenomenology is viable.}
\label{p1rehe}
\end{figure}
Therefore, it is vital that the reheating temperature is
significantly large in order for this combined $R^2$ slow-roll
inflation with a scalar field driven power-law tail inflation to
be phenomenologically viable. In fact, a low-reheating temperature
seems not to be viable at all, since the slow-roll era is
significantly shortened, and in fact the modes corresponding to
the power-law tail might then contribute to the CMB spectrum. This
scenario however is something undesirable. Thus a large reheating
temperature is needed in order for the combined $R^2$ inflation
with a power-law inflationary tail to be a viable inflationary
model. In order to have a good grasp of the phenomenology of the
model under study, let us compare the results with the Planck
data. We shall also consider the case $\beta=0.9$, in addition to
$\beta=0.99$ in order to see how does the EoS parameter $w$ and
the reheating temperature affect the viability of the model. In
Tables \ref{tablerehe1} and \ref{tablerehe2} we gathered the
values of the maximum $e$-foldings number, and the inflationary
observational indices for $\beta=0.99$, and $w=-0.337793$ and for
for $\beta=0.9$, and $w=-0.37931$ respectively. We used three
reheating temperatures $T_r$ a high reheating temperature
$T_r=10^{12}$GeV, an intermediate $T_r=10^{7}$GeV, and a low
reheating temperature, $T_r=10^{2}$GeV. Now in order to assess the
viability of the models, in Figs. \ref{p1rehe} and \ref{p2rehe} we
confront the models with the Planck data for the two distinct
values of $\beta$, namely $\beta=0.9$ and $\beta=0.99$.
Specifically, in Fig. \ref{p1rehe} we present the inflationary
phenomenology for $\beta=0.9$, and $w=-0.37931$ versus the Planck
2018 data. In all the plots of Fig. \ref{p1rehe}, the black thick
line corresponds to the Planck 2018 likelihood curve at $95\%$CL,
the dash dotted curve to the Planck 2018 likelihood curve at
$68\%$CL. In the upper left plot of Fig. \ref{p1rehe}, the red
curve corresponds to $T_r=10^2\,$GeV for $N=[28,32]$, in the upper
right plot of Fig. \ref{p1rehe} the red curve corresponds to
$T_r=10^7\,$GeV for $N=[40,45]$ and in the bottom plot of Fig.
\ref{p1rehe} the red curve corresponds to $T_r=10^{12}\,$GeV for
$N=[50,59]$. As it can be seen, only the high reheating
temperature phenomenology is viable. Also in Fig. \ref{p2rehe} we
present the inflationary phenomenology for $\beta=0.99$, and
$w=-0.337793$ versus the Planck 2018 data. In all the plots of
Fig. \ref{p2rehe}, the black thick line corresponds to the Planck
2018 likelihood curve at $95\%$CL, the dash dotted curve to the
Planck 2018 likelihood curve at $68\%$CL. In the upper left plot
of Fig. \ref{p2rehe}, the red curve corresponds to $T_r=10^2\,$GeV
for $N=[30,36]$, in the upper right plot of Fig. \ref{p2rehe} the
red curve corresponds to $T_r=10^7\,$GeV for $N=[40,48]$ and in
the bottom plot of Fig. \ref{p2rehe} the red curve corresponds to
$T_r=10^{12}\,$GeV for $N=[50,59]$. As it can be seen in this case
too, only the high reheating temperature phenomenology is viable.
Thus the model is viable for a wide choice of EoSs, but only for
high reheating temperatures.

In this section our approach towards describing the inflationary
$R^2$ era combined with the scalar field driven power-law
inflationary tail, was mainly quantitative at the level of field
equations. In the next section we shall approach our description
at the dynamical system level. For generality, we shall allow the
scalar field fluid to interact with the dark matter fluid. As it
proves, and as expected of course, the dark matter fluid plays no
role in the dynamics of the model, even in the presence of an
interaction between the fluids. The result is quite interesting
since we will show that the unstable de Sitter attractor reached
by the $R^2$ model is followed by an inflationary attractor with
$w\leq -1/3$ which is however stable. As we discuss, this
stability is an issue for the combined model and only when the
$R^2$ model fluctuations initiate the reheating era, might make
the model depart from the power-law tail inflationary attractor.

\section{Dynamics of the Universe after Inflation and During the Power-law Inflationary Tail Era}

\begin{figure}[h!]
\centering
\includegraphics[width=20pc]{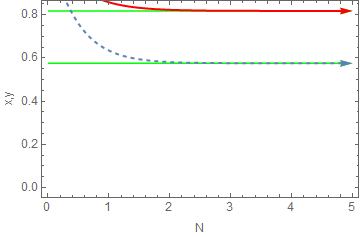}
\includegraphics[width=20pc]{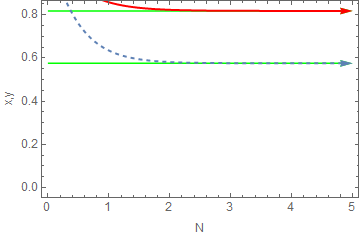}
\includegraphics[width=20pc]{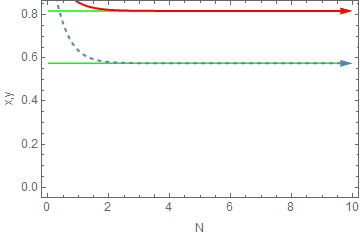}
\includegraphics[width=20pc]{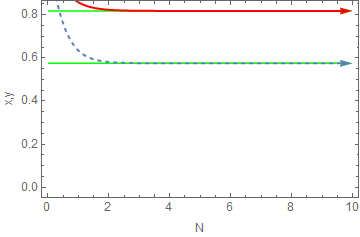}
\caption{Phase space trajectories $x(N)$ (dashed blue curve) and
also $y(N)$ (thick red curve) for the dynamical system
(\ref{dynamicalsystemtwodimensionaldynsubsystemain}) for positive
initial conditions.} \label{plot1}
\end{figure}

\begin{figure}[h!]
\centering
\includegraphics[width=25pc]{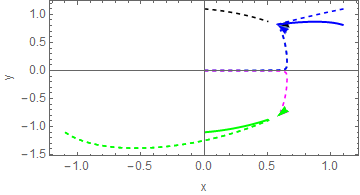}
\caption{Phase space trajectories in the plane $x(N)-y(N)$ plane
for the dynamical system
(\ref{dynamicalsystemtwodimensionaldynsubsystemain}) using various
initial conditions, including negative values for $y(N)$. Notice
the blue thick and the magenta dashed curves, which pass through
the unstable fixed point $P_3^*$ before ending up to the stable
scalar field accelerating attractors $P_4^*$ and $P_5^*$
respectively.} \label{plot2}
\end{figure}

In this section we shall investigate the dynamics of the two
inflationary patches, namely the one driven by the slow-roll $R^2$
gravity and its power-law tail driven by a constant EoS scalar
field, from a dynamical system point of view. Let us start with
the $R^2$ patch first. In the standard $R^2$ gravity, as the
slow-roll inflation era tends to its end, curvature fluctuations
$\langle R^2 \rangle $ become strong enough to make the de Sitter
attractor unstable. This is also the case in our scenario, and as
we now demonstrate, the $R^2$ dynamics is attracted towards an
unstable de-Sitter attractor. This is a unique feature of $R^2$
inflation, which enjoys an elevated importance among inflationary
theories, from a dynamical system point of view, see
\cite{Odintsov:2017tbc}. Considering solely the $F(R)$ gravity
field equations, we introduce the following dimensionless
variables,
\begin{equation}\label{variablesslowdown}
x_1=-\frac{\dot{F_R}(R)}{F_R(R)H},\,\,\,x_2=-\frac{F(R)}{6F_R(R)H^2},\,\,\,x_3=
\frac{R}{6H^2}\, ,
\end{equation}
and by expressing the dynamics of the variables as functions of
the $e$-foldings number, we can write the $F(R)$ gravity field
equations in terms of an autonomous dynamical system as follows,
\begin{align}\label{dynamicalsystemmain}
& \frac{\mathrm{d}x_1}{\mathrm{d}N}=-4-3x_1+2x_3-x_1x_3+x_1^2\, ,
\\ \notag &
\frac{\mathrm{d}x_2}{\mathrm{d}N}=8+m-4x_3+x_2x_1-2x_2x_3+4x_2 \, ,\\
\notag & \frac{\mathrm{d}x_3}{\mathrm{d}N}=-8-m+8x_3-2x_3^2 \, ,
\end{align}
where $m$ is defined to be,
\begin{equation}\label{parameterm}
m=-\frac{\ddot{H}}{H^3}\, .
\end{equation}
The dynamical system of Eq. (\ref{dynamicalsystemmain}) is an
autonomous dynamical system for constant values of the parameter
$m$. For a quasi-de Sitter evolution with scale factor
$a(t)=e^{H_0 t-H_i t^2}$, the parameter $m$ is equal to zero. The
total EoS of the cosmological fluids is equal to \cite{reviews1},
\begin{equation}\label{weffoneeqn}
w_{eff}=-1-\frac{2\dot{H}}{3H^2}\, ,
\end{equation}
which can be expressed in terms of the dimensionless variable
$x_3$ as follows,
\begin{equation}\label{eos1}
w_{eff}=-\frac{1}{3} (2 x_3-1)\, .
\end{equation}
The fixed points of the dynamical system of Eq.
(\ref{dynamicalsystemmain}) with $m=0$, are the following two,
\begin{equation}\label{fixedpointdesitter}
\phi_*^1=(-1,0,2),\,\,\,\phi_*^2=(0,-1,2)\, ,
\end{equation}
with  the corresponding eigenvalues of the linearized matrix
corresponding to the dynamical system for $\phi_*^1$ are $(-1, -1,
0)$, and for fixed point $\phi_*^2$ these are $(1, 0, 0)$. Hence,
the dynamical system of Eq. (\ref{dynamicalsystemmain}) possesses
two non-hyperbolic fixed points, with $\phi_*^1$ being stable and
the fixed point $\phi_*^2$ is unstable. Both fixed points are de
Sitter fixed points, since for both $x_3=2$ and therefore, for
both, the total EoS is $w_{eff}=-1$. The unstable de Sitter fixed
point is more important phenomenologically, since it is directly
related to $R^2$ gravity, and notice that the analysis so far did
not specify an $F(R)$ gravity model. Let us show how this unstable
fixed point is directly related to $R^2$ gravity. To this end,
since $\phi_*^2=(0,-1,2)$ this means that $x_1\simeq 0$ and
$x_2\simeq -1$ which imply,
\begin{align}\label{caseidiffseqns1}
-\frac{\mathrm{d}^2F}{\mathrm{d}R^2}\frac{\dot{R}}{H\frac{\mathrm{d}F}{\mathrm{d}R}}\simeq
0,\,\,\,-\frac{F}{H^2\frac{\mathrm{d}F}{\mathrm{d}R}6}\simeq -1\,
.
\end{align}
Assuming a slow-roll evolution during an inflationary era, the
Ricci scalar curvature is approximated as follows $R\simeq 12
H^2$, for a quasi-de Sitter evolution, thus we can write,
\begin{equation}\label{seconddiff}
F\simeq \frac{\mathrm{d}F}{\mathrm{d}R} \frac{R}{2}\, ,
\end{equation}
which yields,
\begin{equation}\label{approximatersquare}
F(R)\simeq \alpha R^2\, ,
\end{equation}
with $\alpha$ an arbitrary integration constant. Hence, the
unstable de Sitter fixed point of the total de Sitter phase space
of a general $F(R)$ gravity, is always related to an $R^2$
gravity, which is a valuable result. Hence, for $R^2$ gravity, the
dynamical system is attracted towards the unstable quasi-de Sitter
attractor, however once this attractor is reached, the system is
instantly repelled from it. Now, once this instability occurs in
the $R^2$ dynamical system, the power-law inflationary tail
commences, which is governed by a constant EoS scalar field. Let
us assume for the sake of generality that the scalar field fluid
interacts strongly with the dark matter fluid, with an interaction
$Q$ to be specified later. Usually for inflation, the effect of
matter fluids are omitted, since these do not contribute strongly
to the dynamics. As we will show, this is indeed true, since the
effect of the non-trivial interaction on the dynamics is absent
completely. The field equations including the $F(R)$ gravity are,
\begin{align}\label{eqnsofmkotion}
& 3 H^2F_R=\frac{RF_R-F}{2}-3H\dot{F}_R+\kappa^2\left(
\rho_r+\rho_m+\frac{1}{2}\dot{\phi}^2+V(\phi)\right)\, ,\\ \notag
& -2\dot{H}F=\kappa^2\dot{\phi}^2+\ddot{F}_R-H\dot{F}_R
+\frac{4\kappa^2}{3}\rho_r\, ,
\end{align}
\begin{equation}\label{scalareqnofmotion}
\ddot{\phi}+3H\dot{\phi}+V'(\phi)=\frac{Q}{\dot{\phi}}\, ,
\end{equation}
where $Q$ denotes the interaction between the matter and scalar
field fluids. Due to this non-trivial interaction, the dark matter
and scalar field fluids are not perfect fluids as we show shortly.
The field equations can be written in an Einstein-Hilbert form as
follows,
\begin{align}\label{flat}
& 3H^2=\kappa^2\rho_{tot}\, ,\\ \notag &
-2\dot{H}=\kappa^2(\rho_{tot}+P_{tot})\, ,
\end{align}
where $\rho_{tot}=\rho_{\phi}+\rho_G+\rho_r+\rho_m$ denotes the
total energy density, which is composed by all the cosmological
fluids present, and $P_{tot}=P_r+P_{\phi}+P_{G}$ denotes the total
pressure of the cosmological fluids present, which are the dark
matter fluid with $\rho_m$ and with a zero pressure, the scalar
field fluid, with $\rho_{\phi}$ and with pressure $P_{\phi}$,
which are equal to
\begin{equation}\label{rhoscalarandP}
\rho_{\phi}=\frac{\dot{\phi}^2}{2}+V(\phi)\, ,\,\,\,
P_{\phi}=\frac{\dot{\phi}^2}{2}-V(\phi)\, .
\end{equation}
Also the radiation fluid is present with $\rho_r$ and pressure
$P_r=\frac{\rho_r}{3}$, and finally the geometric fluid with
$\rho_{G}$ and pressure $P_G$ which are equal to,
\begin{equation}\label{degeometricfluid}
\rho_{G}=\frac{F_R R-F}{2}+3H^2(1-F_R)-3H\dot{F}_R\, ,
\end{equation}
\begin{equation}\label{pressuregeometry}
P_G=\ddot{F}_R-H\dot{F}_R+2\dot{H}(F_R-1)-\rho_G\, .
\end{equation}
Now the evolution of the fluids is as follows,
\begin{align}\label{fluidcontinuityequations}
& \dot{\rho}_m+3H(\rho_m)=-Q\, , \\ \notag &
\dot{\rho}_{\phi}+3H(\rho_{\phi}+P_{\phi})=Q\, ,
\\ \notag &
 \dot{\rho}_r+3H(\rho_r+P_r)=0\, , \\
\notag & \dot{\rho}_G+3H(\rho_G+P_G)=0\, .
\end{align}
All the fluids compose the total fluid, which is a perfect fluid,
\begin{align}\label{fluidcontinuityequations}
\dot{\rho}_{tot}+3H(\rho_{tot}+P_{tot})=0\, .
\end{align}
Let us assume that the interaction $Q$ has the following form
\cite{Wetterich:1994bg},
\begin{equation}\label{interactionterm1}
Q=\sqrt{\frac{2}{3}}\kappa \beta \rho_m\dot{\phi}\, ,
\end{equation}
with $\beta$ defined earlier, it is the parameter that enters in
the dynamics of the scalar field. Near the end of the slow-roll
era, before the reheating, the constant EoS scalar field controls
the dynamics, so let us assume that also dark matter affects the
dynamics of the system.
\begin{table}[h!]
  \begin{center}
    \caption{\emph{The Fixed Points of the Dynamical System  of Eq. (\ref{dynamicalsystemtwodimensionaldynsubsystemain}) for General Values of $\beta$ and $\lambda$.}}
    \label{table1}
    \begin{tabular}{|r|r|}
     \hline
      \textbf{Name of Fixed Point} & \textbf{Fixed Point Values for General $\beta$ and $\lambda$}  \\
           \hline
      $P_1^*$ & $(x_*,y_*)=(-1,0)$  \\ \hline
      $P_2^*$ & $(x_*,y_*)=(1,0)$\\ \hline
      $P_3^*$ & $(x_*,y_*)=(\frac{2 \beta }{3},0)$ \\ \hline
      $P_4^*$ & $(x_*,y_*)=(\frac{\lambda }{\sqrt{6}},-\frac{\sqrt{2 \beta  \lambda ^2-12 \beta -\sqrt{6} \lambda ^3+6 \sqrt{6} \lambda }}{\sqrt{6} \sqrt{\sqrt{6} \lambda -2 \beta }})$
      \\\hline
      $P_5^*$ & $(x_*,y_*)=(\frac{\lambda }{\sqrt{6}},\frac{\sqrt{2 \beta  \lambda ^2-12 \beta -\sqrt{6} \lambda ^3+6 \sqrt{6} \lambda }}{\sqrt{6} \sqrt{\sqrt{6} \lambda -2 \beta }})$
      \\\hline
      $P_6^*$ & $(x_*,y_*)=(-\frac{3}{2 \beta -\sqrt{6} \lambda },-\frac{\sqrt{\frac{6 \lambda ^2}{2 \beta -\sqrt{6} \lambda }-\frac{2 \sqrt{6} \beta  \lambda }{2 \beta -\sqrt{6} \lambda }-\frac{18}{2 \beta -\sqrt{6} \lambda }-4 \beta +\sqrt{6} \lambda }}{\sqrt{2} \sqrt{\sqrt{6} \lambda -2 \beta }})$
      \\\hline
      $P_7^*$ & $(x_*,y_*)=(-\frac{3}{2 \beta -\sqrt{6} \lambda },\frac{\sqrt{\frac{6 \lambda ^2}{2 \beta -\sqrt{6} \lambda }-\frac{2 \sqrt{6} \beta  \lambda }{2 \beta -\sqrt{6} \lambda }-\frac{18}{2 \beta -\sqrt{6} \lambda }-4 \beta +\sqrt{6} \lambda }}{\sqrt{2} \sqrt{\sqrt{6} \lambda -2 \beta }})$
      \\\hline
    \end{tabular}
  \end{center}
\end{table}
Disregarding the effects of $F(R)$ gravity, the field equations
read,
\begin{equation}\label{subsystemfriedmann}
3H^2=\kappa^2\rho_m+\frac{\kappa^2\dot{\phi}^2}{2}+V\, ,
\end{equation}
which can be cast as,
\begin{equation}\label{friedmannconstraint}
\Omega_m+\Omega_{\phi}=1\, ,
\end{equation}
with,
\begin{equation}\label{omegaphiomegam}
\Omega_{\phi}=\frac{\kappa^2\rho_{\phi}}{3H^2},\,\,\,\Omega_m=\frac{\kappa^2\rho_{m}}{3H^2}\,
.
\end{equation}
The total EoS parameter $w_{tot}$ for the case at hand is,
\begin{equation}\label{totaleosparameter}
w_{tot}=\frac{P_{\phi}}{\rho_{\phi}+\rho_m}=w_{\phi}\Omega_{\phi}\,
,
\end{equation}
and also the total energy density of the scalar field-dark matter
system, satisfies,
\begin{equation}\label{continuitytotal}
\dot{\rho}_{tot}+3H(1+w_{tot})\rho_{tot}=0\, ,
\end{equation}
and also the components satisfy,
\begin{align}\label{fluidcontinuityequations1}
& \dot{\rho}_m+3H \rho_m=-Q\, , \\ \notag &
\dot{\rho}_{\phi}+3H(\rho_{\phi}+P_{\phi})=Q\, .
\end{align}
\begin{table}[h!]
  \begin{center}
    \caption{\emph{Fixed Points of the Dynamical System of Eq. (\ref{dynamicalsystemtwodimensionaldynsubsystemain}) for $\beta=0.99$.}}
    \label{table2}
    \begin{tabular}{|r|r|}
     \hline
      \textbf{Name of Fixed Point} & \textbf{Fixed Point Values for $\beta=0.99$}  \\
           \hline
      $P_1^*$ & $(x_*,y_*)=(-1,0)$  \\\hline
      $P_2^*$ & $(x_*,y_*)=(1,0)$\\\hline
      $P_3^*$ & $(x_*,y_*)=(0.66,0)$ \\\hline
      $P_4^*$ & $(x_*,y_*)=(0.575416,-0.817861)$ \\\hline
      $P_5^*$ & $(x_*,y_*)=(0.575416,0.817861)$ \\\hline
      $P_6^*$ & $(x_*,y_*)=(2.03736,-1.67516)$ \\\hline
      $P_7^*$ & $(x_*,y_*)=(2.03736,1.67516)$ \\\hline
    \end{tabular}
  \end{center}
\end{table}
Let us introduce the following dimensionless variables,
\begin{equation}\label{variablesdynamicalsystem}
x=\frac{\kappa^2\dot{\phi}^2}{6H^2},\,\,\,y=\frac{\kappa^2V}{3H^2}\,
,
\end{equation}
and using these we can express,
\begin{equation}\label{constraintsnew}
w_{\phi}=\frac{x^2-y^2}{x^2+y^2},\,\,\,\Omega_{\phi}=x^2+y^2\leq
1\, ,
\end{equation}
and also the Raychaudhuri equation can be written as follows,
\begin{equation}\label{ryachaudhuru}
-2\dot{H}=3H^2(1+x^2-y^2)\, .
\end{equation}
Using the above,  we can construct the following two-dimensional
autonomous dynamical system \cite{Boehmer:2008av},
\begin{align}\label{dynamicalsystemtwodimensionaldynsubsystemain}
& \frac{\mathrm{d}x}{\mathrm{d}N}=-3x+\frac{\lambda \sqrt{6}}{2}y^2+\frac{3 x}{2}\left(1+x^2-y^2 \right)+\beta \left(1-x^2-y^2\right)\, ,              \\
\notag & \frac{\mathrm{d}x}{\mathrm{d}N}=-\frac{\lambda
\sqrt{6}}{2}x\,y+\frac{3 y}{2}\left(1+x^2-y^2 \right)\, ,
\end{align}
with $e$-foldings number $N$ being the dynamical variable. We
shall study the above dynamical system for $\beta=0.99$ and recall
that $\lambda$ is defined in Eq. (\ref{lambda}).
\begin{table}[h!]
  \begin{center}
    \caption{\emph{Eigenvalues of the Jacobian matrix for the dynamical system (\ref{dynamicalsystemtwodimensionaldynsubsystemain}) for $\beta=0.99$.}}
    \label{table3}
    \begin{tabular}{|r|r|r|}
     \hline
      \textbf{Name of Fixed Point} & \textbf{Eigenvalues} & \textbf{Stability}  \\
           \hline
      $P_1^*$ & $(4.98, 4.72625)$ & unstable
      \\\hline
      $P_2^*$ & $(1.27375, 1.02)$ &
      unstable\\\hline
      $P_3^*$ & $(1.01408, -0.8466)$  &
      saddle\\\hline
      $P_4^*$ & $(-2.1527, -2.00669)$ & stable\\ \hline
      $P_5^*$ & $(-2.1527, -2.00669)$ & stable \\ \hline
      $P_6^*$ & $(4.52339, -4.0064)$ & saddle \\ \hline
      $P_7^*$ & $(4.52339, -4.0064)$ & unstable\\ \hline
    \end{tabular}
  \end{center}
\end{table}
Let us present the fixed points of the dynamical system at hand,
for $\beta=0.99$ and for the corresponding $\lambda$. For general
values of $\beta$ and $\lambda$, the fixed points are given in
\ref{table1}, while for $\beta=0.99$ the fixed points are
presented in Table \ref{table2}. The stability can be determined
by the eigenvalues of the corresponding  Jacobian matrix and these
are given in Table \ref{table3}. Only the fixed points $P_4^*$ and
$P_5^*$ are stable, the fixed points $P_6^*$ and $P_7^*$ are
unphysical. The fixed points $P_4^*$ and $P_5^*$ are stable
attractor points with, totaly scalar field dominated with
$w=-0.337793$. These points are exactly the attractors described
by the power-law tail inflation. As we can see, the interaction
with matter plays no role at all. The fixed points $P_1^*$ and
$P_2^*$ are unstable kinetic attractors, and their significance,
if any, will be determined by a phase space plot. The same
numerical analysis will reveal the attractor property of the fixed
points $P_4^*$ and $P_5^*$. We solved numerically the dynamical
system (\ref{dynamicalsystemtwodimensionaldynsubsystemain}) for a
plethora of initial conditions, and in Fig. \ref{plot1} we present
the $(x(N),y(N))$ trajectories as functions of the $e$-foldings
number $N$. With blue dashed curves we represent $x(N)$, while the
red thick curves represent $y(N)$. The two green lines represent
the points $(0.575416,0.817861)$. As it can be seen, the fixed
point $P_4^*$ is a stable attractor of the theory, but note that
we chose initial conditions to be positive, with regards to
$y(N)$.
\begin{table}[h!]
  \begin{center}
    \caption{\emph{Physical parameters for the fixed points of the dynamical system of Eq. (\ref{dynamicalsystemtwodimensionaldynsubsystemain}) for $\beta=0.99$.}}
    \label{table4}
    \begin{tabular}{|r|r|r|r|r|r|}
     \hline
      \textbf{Name of Fixed Point} & $w_{tot}$ & $\Omega_{\phi}$ & $w_{\phi}$ & $\Omega_m$ &  \textbf{Stability}
      \\ \hline
      $P_1^*$ & 1 & 1 & 1 & 0 & unstable \\ \hline
      $P_2^*$ & 1 & 1 & 1 & 0 & unstable\\ \hline
      $P_3^*$ & 0.4356  & 0.4356 & 1 & 0.5644 & unstable\\ \hline
      $P_4^*$ & -0.337793  & 1 & -0.337793 & 0 & stable\\ \hline
      $P_5^*$ &-0.337793  & 1 & -0.337793 & 0 & stable\\ \hline
      $P_6^*$ & 1.34466 & 6.95699 & 0.193281 & -5.95699 & unstable \\\hline
      $P_7^*$ & 1.34466 & 6.95699 & 0.193281 & -5.95699 & unstable\\
      \hline
    \end{tabular}
  \end{center}
\end{table}
In order to have a more general idea for the behavior of the
trajectories, in Fig. \ref{plot2} we present the phase space
trajectories for the dynamical system
(\ref{dynamicalsystemtwodimensionaldynsubsystemain}) in the
combined $x(N)-y(N)$ plane using various initial conditions,
including negative values for $y(N)$. We can see in Fig.
\ref{plot2} that the stable fixed points $P_4^*$ and $P_5^*$ are
indeed the final attractors and that these are actually scalar
field attractors with a constant EoS $w=-0.337793$. We can also
see an interesting feature, that some of the trajectories pass
from the stiff unstable fixed point $P_3^*$, and this is rather
interesting.
Thus we verified numerically, that the dynamics of inflation after
the $R^2$ slow-roll era is dominated by the constant EoS scalar
field and that the attractor is a final acceleration point which
is scalar field dominated with $w=-0.337793$. As a future study we
leave the combined study of the total $F(R)$ gravity and scalar
field dynamical system. This study however is too extended to be
inserted here and it would be out of the scopes of this article.
We hope though to address this issue in a future work.

\section{Conclusions and Discussion}

In this paper we studied the scenario in which the slow-roll era
of inflation is followed by a power-law inflationary tail governed
by a scalar field with constant EoS parameter. The slow-roll
inflationary era is governed by an $R^2$ gravity. This scenario is
motivated by the fact that the first quantum corrections of a
scalar field in its vacuum configuration contains $R^2$ terms but
also higher order curvature terms. As we showed, there are two
inflationary patches in this framework, one in which the slow-roll
era occurs and the main problems of Standard Big Bang cosmology
are smoothly resolved, and one patch controlled by the scalar
field with constant EoS. The first patch, which is controlled by
the $R^2$ gravity solely, is responsible for the primordial tensor
and scalar perturbations which have large wavelength and exited
the Hubble horizon almost instantly after inflation started. These
modes are basically related with the CMB observations, since these
have wavelength significantly larger than 10Mpc and thus
contribute to the linear modes of perturbations which are observed
by the Planck collaboration. On the contrary, the second
inflationary patch initiates when small wavelength modes enter the
Hubble horizon, and these modes have quite small wavelength, much
smaller than 10Mpc. Thus these modes, although contribute to
structure formation via their tensor and mainly scalar curvature
perturbations generated during this era, cannot have a direct
effect on the CMB. These modes would possibly contribute to the
non-linear features of the CMB, although marginally since the
wavelength of these modes is significantly small, so small that
these are the first modes that reenter the Hubble horizon after
the second inflationary era ends. The functionality of the
secondary inflationary era, namely that of the power-law tail of
the standard slow-roll $R^2$ inflationary era, is that it
basically helps to resolve basic problems of the standard
slow-roll inflationary era, namely it provides an elegant solution
to the TCC problem and also helps resolve the de-Sitter swampland
criterion. We demonstrated explicitly how this resolution may be
achieved, although our crucial assumption is that the radiation
era commences directly after the end of the second inflationary
patch. Another effect which we discussed is the fact that due to
the fact that in the end of first slow-roll inflationary era, the
total background EoS is dominated by the scalar field EoS
parameter, and is thus different from a radiation domination era,
the actual duration of the slow-roll inflationary era is somewhat
shortened, and in fact this feature depends on the total reheating
temperature achieved in the Universe after the two inflationary
eras end. We also studied the phase space of the combined dark
matter-scalar field subsystem for the power-law inflationary patch
and after. As we demonstrated, the final attractors of the
dynamical system are two accelerating attractors which are stable
fixed points of the dynamical system, and as expected, dark matter
does not affect at all the dynamics of the cosmological system.
What we did not address in this article is the study of the total
dynamical system composed by $F(R)$ gravity and the scalar field,
which we aim to do so in a future work. Also, relaxing the TCC
constraints if a post-inflationary reheating era with EoS distinct
than that of radiation must also be appropriately addressed, but
we need to note that this is technically demanding. Also it is
always interesting to discover quantum gravitational effects in
effective actions, except from the standard scalar field $R^2$
effects which we studied here, in the line of research of Refs.
\cite{Miao:2024shs,Miao:2024nsz,Miao:2024atw}. Certainly such
studies are interesting and we hope to address these in some
future works.

\section*{Acknowledgements}

This work was partially supported by the program Unidad de
Excelencia Maria de Maeztu CEX2020-001058-M, Spain (S.D.O). This
research has been is funded by the Committee of Science of the
Ministry of Education and Science of the Republic of Kazakhstan
(Grant No. AP19674478) (Vasilis K. Oikonomou).

\end{document}